\documentclass[aps,prl,twocolumn,showpacs,superscriptaddress,groupedaddress]{revtex4-2}

\usepackage{graphicx}
\usepackage{dcolumn}
\usepackage{bm}

\usepackage[caption=false,font=footnotesize,captionskip=-3mm,farskip=-0mm]{subfig}
\usepackage{amssymb}
\usepackage{epsfig}
\usepackage{amsmath}
\usepackage{textcomp}
\usepackage{url}
\usepackage{float}
\usepackage{color}
\usepackage{cases}
\usepackage{relsize}
\usepackage[normalem]{ulem}
\usepackage{comment}

\newcommand{\black}{\textcolor{black}}
\usepackage{amsmath}
\usepackage{graphicx}
\usepackage{setspace}
\usepackage[colorlinks=true, allcolors=blue]{hyperref}

\newcommand{\Lap}{\mathcal{L}}
\newcommand{\Act}{\mathcal{S}}
\newcommand{\Ham}{\mathcal{H}}
\newcommand{\Ei}[1]{E_{#1}}
\newcommand{\Mem}[1]{M_{#1}}
\newcommand{\Memlap}[1]{\tilde{M}_{#1}}
\newcommand{\Memf}[1]{\mathcal{M}_{#1}}

\begin{document}
\title{Large deviations in non-Markovian stochastic epidemics}
\author{Matan Shmunik$^{a}$} 
\author{Michael Assaf$^{a}$}
\affiliation{$^a$Racah Institute of Physics, The Hebrew University of Jerusalem, Jerusalem 91904, Israel}


\begin{abstract}
We develop a framework for non-Markovian, well-mixed SIR and SIS models beyond mean field, utilizing the continuous-time random walk formalism. 
Using a gamma distribution for the infection and recovery inter-event times as a test case, we derive asymptotical late-time master equations with effective memory kernels and obtain analytical predictions
for the final outbreak size distribution in the SIR model, and quasistationary distribution and disease lifetime in the SIS model. 
We show that varying the width of the inter-event time distribution can greatly alter the outbreak size distribution or the disease lifetime. We also show that rescaled Markovian models may fail to capture fluctuations in the non-Markovian case. \textcolor{black}{Overall, our analysis, confirmed against numerical simulations,  paves the way for studying large deviations in structured populations on degree-heterogeneous networks.}
\end{abstract}
\maketitle

\textit{Introduction. }
The challenge of modeling and predicting pandemics has motivated extensive development of deterministic and stochastic epidemiological models~\cite{anderson1991infectious, kermack1927contribution, allen1994some, mollison1995epidemic, daley1999epidemic, hethcote2000mathematics, keeling2005networks, neipel2020power}. Their importance was underscored during COVID-19~\cite{yang2021rational, ihme2021modeling}, when forecasts guided policies worldwide. These models typically divide populations into disease-status compartments. The SIS model~\cite{hethcote1989three} with susceptible and infected compartments, describes infections with short-lived immunity (e.g., influenza, or the common cold), while the SIR model~\cite{neipel2020power} adds a recovered compartment to capture long-term immunity, which is relevant for diseases such as measles, smallpox, polio and COVID-19.
Generalized models include reinfection (SIRS), an exposed compartment (SEIR)~\cite{hethcote2000mathematics}, time-varying rates~\cite{aron1984seasonality,shaw2008fluctuating}, or structured populations, where network-based approaches~\cite{pastor2015epidemic, karrer2010message, lindquist2011effective, miller2012edge, keeling1999effects, house2011insights,cai2016solving} capture heterogeneity in population structure.
Mean-field quantities have been computed for both models: the mean outbreak size and epidemic threshold in the SIR model, and  endemic state and relaxation dynamics in the SIS model~\cite{hethcote1989three, daley1999epidemic, allen1994some, pastor2015epidemic}. Stochastic dynamics have also been explored, including the outbreak-size distribution for the SIR model and quasistationary distribution (QSD) and disease mean time to extinction (MTE)  for the SIS model~\cite{ovaskainen2001quasistationary,dykman2008disease,shaw2008fluctuating,ovaskainen2010stochastic,assaf2010extinction,ovaskainen2010stochastic,hindes2016epidemic,assaf2017wkb,hindes2019degree,hindes2022outbreak,hindes2023fluctuating,korngut2025weighted,korngut2025impact}. 

A key limitation of most approaches is the Markovian assumption~\cite{keeling2005networks, volz2008sir, sherborne2018mean}\black{, which states that the system dynamics depend solely on the present state, and are independent of prior history. Under this assumption, both infection and recovery periods are exponentially distributed. However, empirical evidence suggests that transmission and recovery rates are often explicitly time-dependent, and consequently, the corresponding waiting-time (WT) distributions are more accurately described by non-exponential distributions, such as log-normal, gamma, or Weibull~\cite{bailey1954statistical, gough1977estimation, wearing2005appropriate}. As a result, infection and recovery dynamics are generally non-Markovian, rendering analytical frameworks based on Markovian assumptions often inadequate for describing such processes.}

\black{Most studies investigating non-Markovian epidemic processes on various network topologies, have focused on the mean-field aspects of the dynamics~\cite{cator2013susceptible,van2013non,boguna2014simulating,kiss2015generalization,rost2015impact,kiss2017non,sherborne2018mean,rost2018pairwise,masuda2018gillespie,starnini2017equivalence}. These studies have shown that, even when average rates are unchanged, non-exponential infection and recovery times can drastically affect disease spread, prevalence, and overall epidemic bahavior. Conversely, certain features of non-Markovian dynamics can still be approximated by rescaling Markovian rates, see, e.g., Refs.~\cite{starnini2017equivalence,feng2019equivalence}}. In this context,  Boguñá \textit{et al.}~\cite{boguna2014simulating} proposed a generalized Gillespie algorithm applicable to non-exponential infection and recovery processes, which was later refined into the more efficient Laplace Gillespie method by Masuda \textit{et al.}~\cite{masuda2018gillespie}. Notably, while the role of noise in non-Markovian epidemics has also been studied in both SIR~\cite{ball1986unified,startsev1997distribution,clancy2014sir,wilkinson2018impact} and SIS~\cite{cator2013susceptible} models, a systematic analysis of the interplay between non-Markovianity and demographic noise on large deviations in these models, remains lacking.

Recent progress has extended non-Markovian frameworks to chemical and biological systems, using the continuous-time random walk  formalism with non-exponential WTs~\cite{aquino2017chemical}. This was later adapted in~\cite{vilk2024gene, vilk2024escape, vilk2025rock} to models of gene expression, population dynamics, and competition. 
Here, we combine this framework with the WKB approach~\cite{dykman1994large, ovaskainen2010stochastic, assaf2017wkb} of Hindes~\textit{et al.}~\cite{hindes2022outbreak}  to study non-Markovian epidemic dynamics beyond mean field in a well-mixed setting. We calculate the memory kernels from general WTs and derive the effective late-time master equation~\cite{aquino2017chemical, vilk2024escape}. We then compute the mean outbreak size and its full distribution within the SIR model, using gamma-distributed WTs as a prototypical example. We then analyze the SIS model, deriving the metastable mean, its full distribution (QSD), and MTE---the first passage time to the absorbing, disease-free state~\cite{redner2001guide,assaf2010extinction}. Finally, we test our formalism using empirical WTs for both infection and recovery.

\black{It is important to note that, in this work we focus on well-mixed populations, i.e., fully-connected networks, in which all individuals are considered identical. As such, for simplicity, we consider only two global reactions: infection and recovery, both with generic WT distributions, whose mean depends on the system's state. While this constitutes an oversimplified representation of realistic scenarios, it turns out that our model can reproduce, at least qualitatively,  key features of large deviations in non-Markovian epidemic dynamics on complex networks.} 

\textit{Non-Markovian SIR model. } 
Within the well-mixed SIR model, given that the rates of infection and recovery per individual  are $\beta$ and $\gamma$, respectively, the  discrete-state stochastic reactions can be written as
        $(S, I) \!\to\! (S \!-\! 1, I \!+\! 1)$ with mean rate $\beta S I/N$, and $(I, R) \!\to\! (I \!-\! 1, R \!+\! 1)$ with mean rate $\gamma I$.
Defining the population fractions as $x_s = S/N$, $x_i = I/N$ and $x_r=R/N$, assuming $N\gg1$ and ignoring  noise, the mean-field rate equations read
\begin{equation}
       \dot{x}_s=-\beta x_s x_i,\quad \dot{x}_i=\beta x_s x_i-\gamma x_r, \quad\dot{x}_r=\gamma x_i,
    \label{eq:rate_equations}
\end{equation}
such that $x_s\!+\!x_i\!+\!x_r\!=\!1$. Here the underlying assumption is that  reactions occur with exponential WTs. Yet,
in reality, the WT distributions denoted by $\psi_1(t)$ for infection and $\psi_2(t)$ for recovery, are not necessarily exponential.

Initially we focus on the case of \black{non-Markovian, gamma-distributed infection and Markovian recovery~\cite{van2013non,boguna2014simulating}}, while other WT choices are discussed in the Supplementary Information (SI). The WT distributions satisfy
\begin{equation}
    \label{phipsi}\begin{gathered}
    \hspace{-3mm}\psi_1(t) = (\alpha \lambda_1)^\alpha t^{\alpha-1}e^{-\alpha \lambda_1 t}/\Gamma(\alpha),\quad \psi_2(t) = \lambda_2 e^{-\lambda_2 t},
    \end{gathered}
\end{equation}
where \black{$\Gamma\left(z\right)=\int_0^\infty t^{z-1}e^{-t}dt\,$ is the gamma function.} Notably, the \black{definition of the gamma distribution here is such that the shape parameter $\alpha$ does not affect the mean.} The WTs means are set to equal the Markovian counterparts, $\lambda_1=N R_0 x_s x_i$ and $\lambda_2=N x_i$ respectively,  where $R_0=\beta/\gamma$ is the basic reproductive number. \black{The shape parameter $\alpha$ controls the tail of the WT distribution, and as such, can be effectively viewed as a form of memory, where lower values of $\alpha$ entail stronger memory (broader distribution) while higher values of $\alpha$ entail weaker memory (narrower distribution).}

To determine the outbreak size distribution within the SIR model, we write the corresponding non-Markovian master equation for the probability $P_{S,I}$ to find $S$ susceptibles and $I$ infected at time $t$, which reads~\cite{vilk2024gene,vilk2024escape}
\begin{eqnarray}\label{masterSIR}
    && \hspace{-3mm}\frac{\partial P_{S,I}}{ \partial t}=\int_{0}^{t}\left[ \Mem{1}(S+1,I-1,t-t')P_{S+1,I-1}(t')\right.\nonumber\\
    &&\hspace{-3mm}\left.-\Mem{1}(S,I,t-t')P_{S,I}(t')+\Mem{2}(S,I+1,t-t')P_ {S,I+1}(t')\right.\nonumber\\
    &&\hspace{-3mm}\left.-\Mem{2}(S,I,t-t')P_{S,I}(t')\right]\,dt'.
\end{eqnarray}
In both infection and recovery, the dynamics depend on the full time history via the memory kernels $\Mem{i}$. The latter can be found by Laplace-transforming the master equation and using the specific structure of the WT distributions, $\psi_i(t)$, see SI, Sec.~A.  Even though the SIR dynamics has no metastability, and includes an infectious wave which quickly dies out, to explore the final outbreak distribution it suffices to look at the asymptotic late-time dependence of $\Mem{i}$ in the limit of $t\to \infty$~\cite{turkyilmazoglu2021explicit}~\footnote{This is justified since for $N\gg1$ the epidemic duration is still much longer than the system's  relaxation time~\textcolor{blue}{[57]}.}. Below we will also provide numerical justification for this claim. Thus, 
we use the final value theorem: $\lim_{t\to\infty} f(t)=\lim_{u\to0} u \tilde{f}(u)$, for the memory kernel functions, where $u$ is the Laplace variable. 
Defining the normalized asymptotic memory kernels: $
\Memf{i}:=(1/N)\lim_{t\to\infty} \Mem{i}(t)=(1/N)\lim_{u\to0} \Memlap{i}(u)$ (see details in the SI, Sec.~A) and using~(\ref{masterSIR}), we obtain the effective late-time master equation for the SIR model
\begin{eqnarray}
    \frac{\partial P_{S,I}}{\partial t}&=&N\left[\Memf{1}(S+1,I-1)P_{S+1,I-1}
    -\Memf{1}(S,I)P_{S,I}\right.\nonumber\\
    &+&\left.\Memf{2}(S,I+1)P_{S,I+1}-\Memf{2}(S,I)P_{S,I}\right],\\
    \label{effectiveSIR}    &&\hspace{-14mm}\text{with}\quad\Memf{1}=\frac{x_i}{\left[1+1/(R_0x_s\alpha)\right]^{\alpha}-1},\quad \Memf{2}=x_i,
    \label{eq:analytical_memf}
\end{eqnarray}
where $\Memf{1}$ and $\Memf{2}$, are the effective infection and recovery rates under \black{non-Markovian (gamma-distributed) infection and Markovian recovery. For $\alpha=1$ (exponential infection), we recover the Markovian rate} $\Memf{1}=R_0 x_s x_i$, whereas decreasing $\alpha$ increases the infection rate and greatly alters the dynamics. Notably, $\Memf{1}$ and $\Memf{2}$ can also be found for other choices of WTs, see SI, Sec.~B.

Having found the effective rates, we now use the WKB approach as in~\cite{hindes2022outbreak} to compute the final outbreak-size distribution. Substituting  $P_{S,I}\sim e^{-N\Act(x_s,x_i,t)}$  into Eq.~(\ref{effectiveSIR}) yields, in the leading order in $N\gg 1$~\footnote{\black{Note that, the leading-order WKB approximation here is valid as long as the total action, $N\mathcal{S}$, is large}~\cite{dykman1994large,assaf2010extinction,assaf2017wkb}.}, a Hamilton-Jacobi equation $\partial \Act(x_s,x_i,t)/\partial t+\Ham(x_s,x_i)\!=\!0$, where $\Act(x_s,x_i,t)$ is the action function, with the Hamiltonian
\begin{equation}
\Ham\equiv \Memf{1}(x_s,x_i)\left(e^{p_i-p_s}\!-\!1\right)\!+\!\Memf{2}(x_s,x_i)\left( e^{-p_i}\!-\!1\right).
\label{eq:hamiltonian}
\end{equation}
Here, we have defined the momenta $p_i=\partial \Act/\partial x_i$, $p_s=\partial \Act/\partial x_s$ in analogy to classical mechanics. In the SI, Sec.~C we show that $p_i$ is a constant of motion, and use Hamilton's equations to derive a relation between $p_i$ and final susceptible fraction $x_s^*\!=\!x_s(t\!\to\!\infty)$. Defining $m=e^{p_i}$, we find an implicit relation between $m$ and $x_s^*$
\begin{eqnarray}
        &\int_1^{x_s^*} [m\left(\Memf{1}/\Memf{2}+1\right)-1]^{-1}dx_s=\int_0^{1-x_s^*} dx_r.
        \label{eq:m_relation}
    \end{eqnarray}
This allows to compute the action function $\Act(x_s,x_i,t)=\int_0^t(p_s\dot{x}_s+p_i\dot{x}_i-\Ham)dt'$~\cite{hindes2022outbreak}
(see SI, Sec.~C), which reads   
\begin{equation}
    \hspace{-2mm}\Act(x_s^*)\!=\!\!\int_1^{x_s^*}\!\!\ln\left\{ m^2\Memf{1}\Big/\left[m\left(\Memf{1}\!+\!\Memf{2}\right)\!-\!\Memf{2} \right]\!\right\}\!dx_s,
        \label{eq:action_sir}
\end{equation}
and is a function of $x_s^*$ only via Eq.~(\ref{eq:m_relation}). Finally, the final outbreak distribution is given by $P(x_s^*)\!\sim \!e^{-N\Act(x_s^*)}$. 

To quantify outbreak variability, we compute the standard deviation by  expanding the action $\Act$ to second order around the mean final susceptible fraction $\bar{x}_s^*$, obtained at $p_s=p_i=0$.
Plugging this in Eq.~(\ref{eq:m_relation}) we get 
$\int_{\bar{x}_s^*}^1 (\Memf{2}/\Memf{1})dx_s\!=\!\int_0^{1-\bar{x}_s^*}\!\! dx_r$,  an implicit equation for  $\bar{x}_s^*$.
Using $\Memf{i}$ from~(\ref{eq:analytical_memf}) yields: \black{$1-\bar{x}_s^*=(2\alpha R_0)^{-1}f(\bar{x}_s^*)$}, with
\begin{equation}
      \black{ f(\bar{x}_s^*)\!=\!\mathrm{B}\!\left(\!\frac{\alpha R_0}{1\!+\!\alpha R_0};\!1\!-\!\alpha,-1\!\right)\!-\!\mathrm{B}\!\left(\!\frac{\alpha R_0 \bar{x}_s^*}{1\!+\!\alpha R_0\bar{x}_s^*}\!;1\!-\!\alpha,-1\!\right)\!,}
\end{equation}
where $\mathrm{B}(x;a,b)\!=\!\!\int_0^x \!t^{1\!-\!a}(1\!-\!t)^{1\!-\!b}dt$ is the incomplete beta function. Here, for $\alpha\!=\!1$, we recover the known mean-field result of $1-\bar{x}_s^*\!=\!-\!\ln{\bar{x}_s^*}/R_0$~\cite{neipel2020power}.

Using  Eq.~(\ref{eq:action_sir}) for the action $\Act$, we can derive the standard deviation $\sigma$ by performing a Gaussian approximation on $P(x_s^*)$ around $\bar{x}_s^*$, yielding $\sigma \!\simeq\! \left|N \Act''(\bar{x}_s^*)\right|^{-1/2}$~\cite{hindes2022outbreak}.
\black{Defining $\xi(\bar{x}_s^*)\!=\!\Memf{2}(\bar{x}_s^*)/\Memf{1}(\bar{x}_s^*)$ and $I(\bar{x}_s^*) = \int_{\bar{x}_s^*}^1\xi(z)^2dz$, we find (see SI, Sec.~C)
\begin{equation}
   \left.\Act''(x_s^*)\right|_{\bar{x}_s^*} =\left[\xi(\bar{x}_s^*)-1\right]^2/\left[1-\bar{x}_s^*+I(\bar{x}_s^*)\right],
\end{equation}
where using Eq.~(\ref{eq:analytical_memf}), $\xi(\bar{x}_s^*)\!=\![1\!+\!1/(R_0\alpha \bar{x}_s^*)]^{\alpha}\!-\!1$,  and $I(\bar{x}_s^*)$ can be expressed by beta functions.}

\begin{figure}[ht]
    \centering
    \includegraphics[width=1\linewidth]{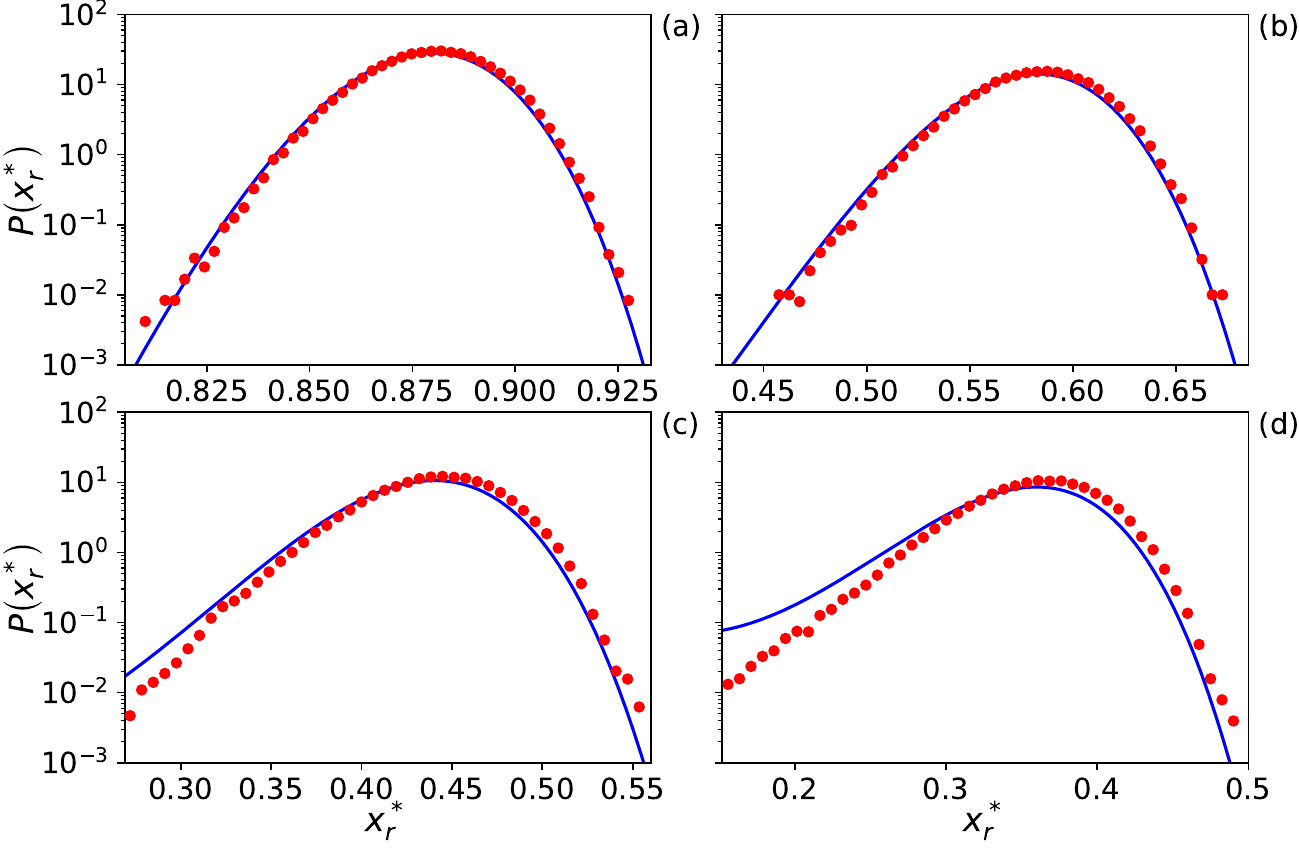}
    \vspace{-9mm}\caption {\black{Outbreak size  distributions for the SIR model, with $\alpha=0.5$, $1$, $1.5$ and $2$ in (a)--(d), for  gamma-distributed infection and Markovian recovery, $N=5000$, $R_0=1.5$, and $10^5$ runs per $\alpha$. Simulations (symbols) are compared with theory (solid line). In all panels $I(0)=1$, and simulations resulting in outbreak fractions  $<10^{-2}$ were omitted, see [61]}.}
    \label{fig1}
\end{figure}

\begin{figure}[ht]
    \centering
    \includegraphics[width=1.02\linewidth]{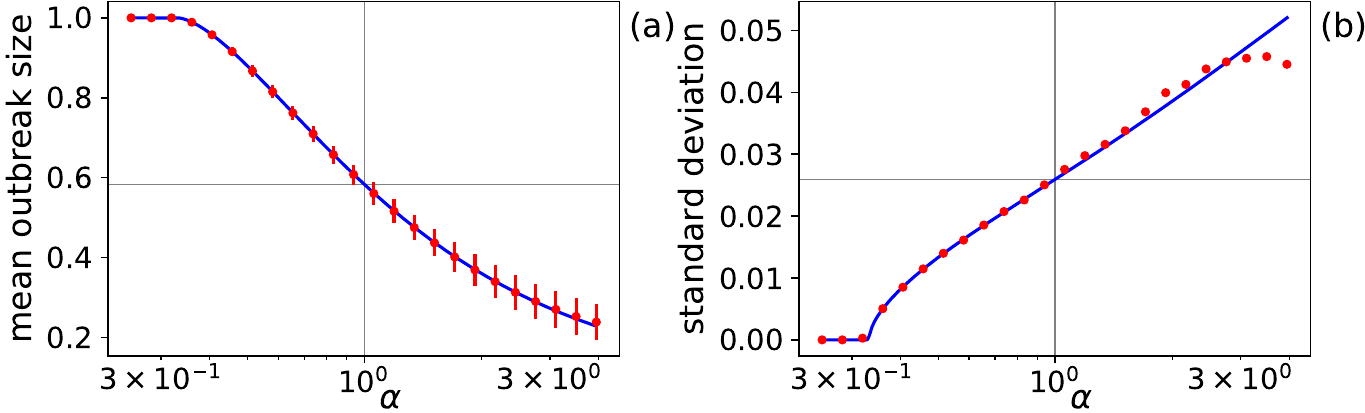}
    \vspace{-8mm}\caption {Normalized mean $x_r^*$ (a) and normalized standard deviation $\sigma$ (b) versus $\alpha$.  Here infection is gamma-distributed and  recovery is exponential, $N=5000$, $R_0=1.5$ and  $10^4$ runs per $\alpha$. Simulations (symbols) are compared with  theory (solid line). In all panels $I(0)=1$, and \black{simulations resulting in} outbreak fractions $< 10^{-2}$ were omitted, see [61].}
    \label{fig2}
\end{figure}

To validate our theory, we ran simulations using the next reaction method with simplified WT sampling~\footnote{\black{In each realization, we sample the WTs for both infection and recovery processes from the given WT distributions, execute the reaction that occurs first, advance the system time accordingly, update the distributions’ mean rates, reset the timers (as done in Ref.~\cite{masuda2018gillespie} when the WT means are state-dependent), and repeat until either disease extinction occurs or the maximum time is reached.}}, see~\cite{anderson2007modified}, which in this case is accurate and faster compared to the algorithms in Refs.~\cite{boguna2014simulating, masuda2018gillespie}.
\black{Figures~\ref{fig1} and \ref{fig2} show results for the non-Markovian  SIR model with gamma-distributed infection and exponential recovery. In Fig.~\ref{fig1} we show examples of the final outbreak size distribution for four different $\alpha$'s~\footnote{\black{In these simulations, we performed a sufficiently large number of realizations so that, even after excluding those with a final outbreak fraction $<10^{-2}$, we still retained the desired number of samples, see Figs.~\ref{fig1} and \ref{fig2}. This allowed us to probe the distribution around the mean without the influence of near-immediate extinction events}.}. Here, theory and simulations agree well as long as the action is large~\cite{dykman1994large,assaf2010extinction,assaf2017wkb,hindes2022outbreak}.
In Fig.~\ref{fig2} we show the mean outbreak fraction and its variance as function of $\alpha$. As $\alpha$ increases, typical infection times becomes longer, and the mean  drops, while the standard deviation increases. We also see that the WKB approximation loses its accuracy as the mean outbreak size approaches zero.}

\textit{Non-Markovian SIS model. }
Here,
a susceptible can get infected at a rate $\beta$ and infected individuals recover at a rate $\gamma$. The Markovian mean-field dynamics satisfy
$\dot{x}_i=\beta x_s x_i-\gamma x_i$,
where $x_i+x_s=1$. Notably, the late-time dynamics of this model are markedly different from the SIR model. Here, for $N\gg 1$, the system enters a long-lived metastable state, which then slowly decays due to an exponentially-small probability flux into the absorbing state at $I=0$. As we are interested in finding the statistics of the metastable state and MTE, we again use the long-time effective memory kernels defined above, and arrive at the late-time master equation
\begin{eqnarray}
   \partial P_{I}/\partial t&=&N\left[\Memf{1}(I-1)P_{I-1}-\Memf{1}(I)P_{I}\right.\nonumber\\
    &+&\left.\Memf{2}(I+1)P_{I+1}-\Memf{2}(I)P_{I}\right].
    \label{eq:effective_master_equation_sis}
\end{eqnarray}
This master equation is valid after the initial relaxation period near the endemic state, $\tau_{r}$, which we compute below, and $\Memf{i}$ are given by Eq.~(\ref{eq:analytical_memf}) with $x_s=1-x_i$.

To quantify the effect of non-Markovian infection on the endemic state $x_i^*$, we multiply Eq.~(\ref{eq:effective_master_equation_sis}) by $I$ and sum over all $I$. 
This yields the modified rate equation $\langle \dot{x}_i\rangle = \Memf{1}(\langle x_i\rangle) - \Memf{2}(\langle x_i\rangle)$, with $x_i^*$ found by solving $\Memf{1}(x_i^*) = \Memf{2}(x_i^*)$. Using~(\ref{eq:analytical_memf}) with $x_s = 1 - x_i$ gives
\begin{equation}\label{meannonmar}
    \black{x_i^*=1-[\alpha R_0(2^{1/\alpha}-1)]^{-1}.}
\end{equation}
At $\alpha=1$ (exponential WTs), we recover the Markovian result $x_i^*=1-1/R_0$. 
However, at $\alpha\neq 1$, one can define a new effective $R_0$, \black{$R_0^{\mathrm{eff}}=\alpha R_0(2^{1/\alpha}-1)$, such that $x_i^*=1-1/R_0^{\mathrm{eff}}$. This means that, in order to reproduce the effect of non-Markovian infection, one can instead take Markovian infection and change the infection rate $\beta\to \beta \alpha(2^{1/\alpha}-1)$ such that at $\alpha\to 0$, the Markovian infection rate should tend to $\infty$, while at $\alpha\to\infty$, the infection rate should tend to $\beta\ln 2$. Consequently, we find that, at $\alpha<1$ an endemic state can persist even when each infected transmits to fewer than one other individual, or in other words, a subcritical epidemic with $R_0<1$ may still reach a nonzero endemic state.}
Note that the relaxation time of the dynamics $\tau_{r}=\left[\Memf{2}{}'\left(x_i^* \right)-\Memf{1}{}'\left(x_i^* \right)\right]^{-1}$ can also be computed.
Using Eq.~(\ref{eq:analytical_memf}) we obtain \black{$\tau_r=2^{1/\alpha-1}/\left\{\alpha\left(2^{1/\alpha}\!-\!1\right)\left[R_0\alpha\left(2^{1/\alpha}\!-\!1\right)\!-\!1\right]\right\}$},
which coincides with the known result of $\tau_{r}=1/(R_0-1)$ at $\alpha=1$.

We now analyze the effective master equation~(\ref{eq:effective_master_equation_sis}) to determine the QSD of the metastable state and the MTE under non-Markovian recovery. Assuming the metastable state decays slowly, we set $P_I(t) = \pi(x_i) e^{-t/\tau_{ext}}$, where $\pi(x_i)$ is the QSD and $\tau_{ext}$ is the exponentially large MTE. We now apply the WKB approximation, $\pi(x_i) \sim e^{-N\Act(x_i)}$, where $\Act(x_i)$ is the action function. Expanding to leading order in $N \!\gg \!1$, and using~(\ref{eq:analytical_memf}) we find $\Act'(x_i) = \ln[\Memf{2}(x_i)/\Memf{1}(x_i)]=\ln\{[1+1/(R_0\alpha(1\!-\!x_i))]^{\alpha}-1\}$, which provides the QSD, $\pi(x_i)$
\begin{equation}\label{qsdsis}
    \pi(x_i)\sim e^{-N\int_{x_i^*}^{x_i}\ln\left[\Memf{2}(x')/\Memf{1}(x')\right] \,dx'}.
\end{equation}
The QSD can be explicitly written using Eq.~(\ref{eq:analytical_memf}) in terms of the elementary functions, for each value of $\alpha$, and generalizes the result for Markovian reactions~\cite{ovaskainen2001quasistationary,assaf2010extinction}. 
In Fig.~S1 we show excellent agreement between simulated QSDs for various values of $\alpha$ and our theoretical prediction, see details in the SI, Sec.~\black{D}. 

\begin{figure}[ht]
    \centering
    \includegraphics[width=1\linewidth]{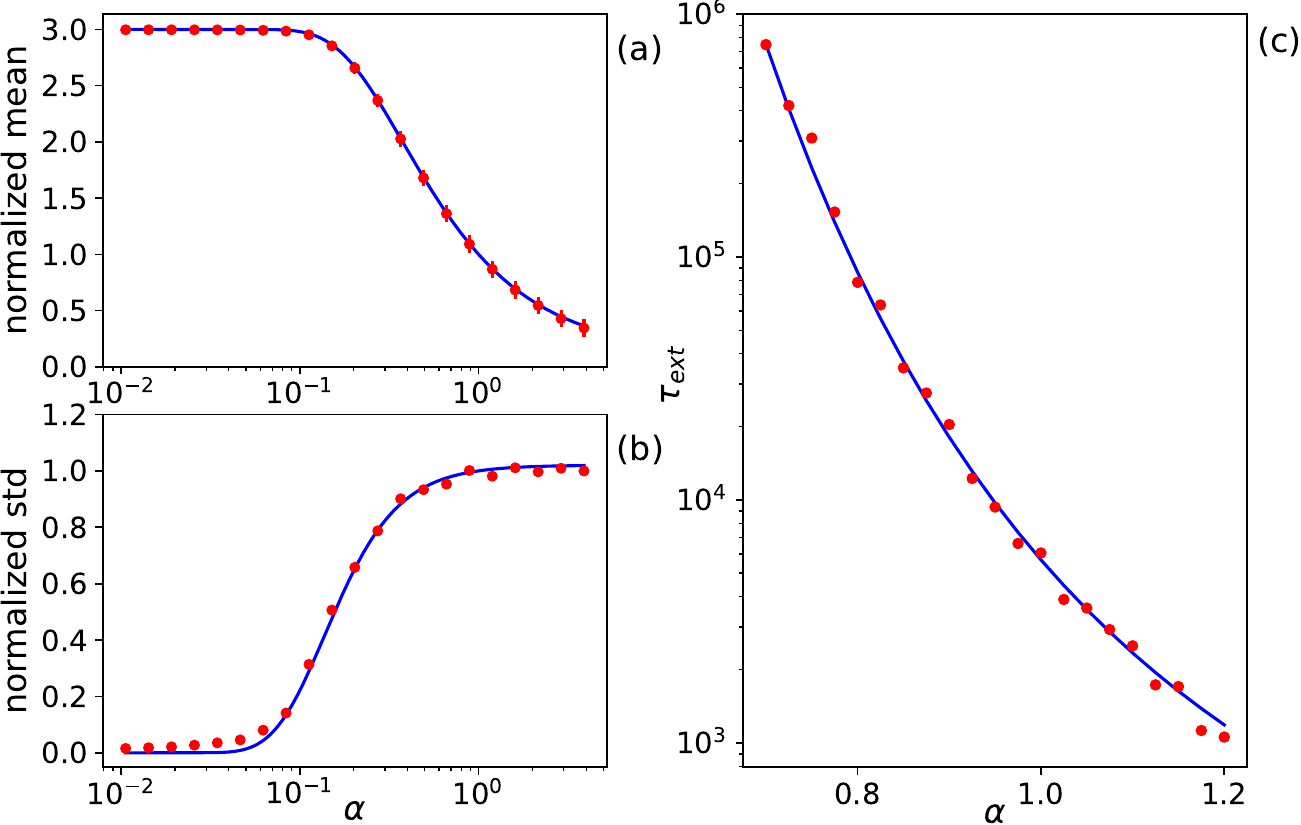}
    \vspace{-8mm}\caption {In (a) and (b) we show the normalized mean $x_i^*$ and normalized standard deviation $\sigma$ versus $\alpha$, for gamma-distributed infection
and Markovian recovery. Here,  $N=1000$, $R_0=1.5$, 
    and we ran $10^3$ realizations per $\alpha$.  Simulations  (symbols) are compared with theory (line). In  (c) we plot the MTE versus $\alpha$  for the same WTs, $N=110$, $R_0=1.51$ 
    and $10^2$ runs per $\alpha$. Simulations (symbols) are compared with  theory (line), see text. In (a-c), $x_i(0)=x_i^*$. 
    }
    \label{fig3}
\end{figure}

We can also compute the QSD's  variance, $\sigma^2$, via a  Gaussian approximation around $x_i^*$~\cite{assaf2010extinction}. This yields
\begin{equation}\label{nonmarsig}
    \black{\sigma^2=\frac{2^{1/\alpha-1}}{N R_0\alpha^2(2^{1/\alpha}-1)^2}}=\frac{1-x_i^*}{2\alpha N (1-2^{-1/\alpha})},
\end{equation}
where we have used Eq.~(\ref{eq:analytical_memf}). At $\alpha=1$, the Markovian result is recovered, $\sigma^2\!=\!1/(NR_0)$. As $\alpha$ increases, typical fluctuations can greatly increase, making disease clearance more likely.
In Fig.~\ref{fig3}(a,b) we show how the shape parameter $\alpha$ affects the metastable mean~(\ref{meannonmar}) and its standard deviation~(\ref{nonmarsig}), both normalized by their Markovian values [$x_i^*(\alpha\!=\!1)\!=\!1/3$, $\sigma(\alpha\!=\!1)\!\simeq\!0.026$ for $R_0=1.5$]. As $\alpha$ increases, the mean decreases drastically~\cite{van2013non,boguna2014simulating}, and the variance increases; this occurs since the typical time between infection events gets longer as the WT distribution becomes narrower. At $\alpha\to 0$, the times between infection events is very short and the mean approaches $1$, while the variance approaches $0$. 

The dependence of the variance on $\alpha$ as observed in Fig.~\ref{fig3}(b) can be explained by looking at the right-hand-side of Eq.~(\ref{nonmarsig}), where the standard deviation is expressed as a function of the mean $x_i^*$ and $\alpha$ (\black{instead of $R_0$ and $\alpha$})~\footnote{Notably, the dependence of $\sigma$ on $x_i^*$ and $\alpha$  remains identical also when infection is exponential and recovery is gamma-distributed, even though $x_i^*$ changes in that case.}.
This reveals two effects shaping $\sigma$. The first and strongest is the monotone relation $\sigma\!\sim\!\sqrt{1\!-\!x_i^*}$, dominating the behavior. The second effect stems from non-Markovianity, $\sigma\!\sim\!\left[2\alpha(1\!-\!2^{-1/\alpha})\right]^{-1/2}$; for  non-Markovian infection, it partially opposes the main effect of increase in $\sigma$ as $\alpha$ is increased; yet, it reinforces the main effect in the case of non-Markovian recovery   (see SI, Sec.~\black{E}). Notably, as $x_i^*$ approaches $0$, the WKB approximation is expected to break down as the action is no longer large. This effect is more evident for non-Markovian recovery, see SI, Sec.~E.

Finally, having computed the complete QSD~(\ref{qsdsis}), we can evaluate the MTE, using $\tau_{ext}\sim e^{N \Act(0)}$, where $\Act(0)$ is the action barrier to extinction~\cite{dykman1994large,assaf2010extinction,assaf2011fixation,assaf2017wkb}. The integral in Eq.~(\ref{qsdsis}) from $x_i^*$ to $0$ yields the MTE, and can be explicitly computed for each $\alpha$. In particular, a simplified expression can be obtained, at $|\alpha-1|\ll 1$, which can provide insight on how the disease lifetime changes as one deviates from Markovianity. In the leading $|\alpha-1|\ll 1$ order, we obtain
\black{$\Act(0)\!\simeq\! {\cal S}_0\!-\!
(\alpha \!-\! 1)/(2 R_0)
\!\!\left[\!(R_0 \!+\! 1)^2 \ln\!\left(\!1 \!+\! R_0^{-1}\!\right)\!-\!
R_0 \!+\! 1 \!-\! \ln (R_0/16)\right]$,
where ${\cal S}_0=\ln R_0+R_0^{-1}-1$ is the action barrier in the $\alpha=1$ case~\cite{ovaskainen2001quasistationary,ovaskainen2010stochastic,assaf2010extinction,assaf2017wkb}.
In Fig.~\ref{fig3}(c) we plot the MTE versus $\alpha$  for gamma-distribued infection and exponential recovery.  Very good agreement holds as long as the action is large (i.e., $\ln\tau_{ext}\gg 1$), which is the case for the entire range of the figure. Here, the solid line shows the theoretical result, $\tau_{ext}
\simeq A(\alpha,R_0,N)e^{N\mathcal{S}(0)}$; it includes an additional preexponential factor $A(\alpha,R_0,N)$~\cite{assaf2010extinction,assaf2017wkb},  numerically fitted (for fixed $N$ and $R_0$) to be $A\sim \alpha^2$.}

\textit{Discussion.} We have studied how non-Markovian reactions shape long-term epidemic dynamics, by deriving the late-time asymptotics of the master equation, and  memory kernels emanating from the WTs.   
\black{We  have shown, within the SIR model, that under non-Markovian infection and exponential recovery, the mean outbreak size and its entire distribution strongly depend on the shape parameter $\alpha$; notably, as $\alpha$ grows, the outbreak risk greatly diminishes. Within the SIS model, we have shown that as $\alpha$ increases, disease prevalence decreases~\cite{van2013non,boguna2014simulating}, and  the risk of disease eradication greatly increases.}

Having focused so far on non-Markovian infection, we now show its applicability to realistic scenarios with both infection and recovery being non-Markovian. Empirical studies of acute infections like influenza and COVID-19 consistently report nonexponential (Gamma, Weibull, or log-normal) WTs for infection and recovery, typically with gamma shape parameters $1<\alpha_\text{inf},\alpha_\text{rec}<6$~\cite{Zhang2020,Bi2020,Linton2020,He2020,Ali2020,Park2020,lee2024variability,carrat2008time,byrne2020inferred,voinsky2020effects}.  Infection WTs typically measure the time from inferred exposure or symptom onset to secondary transmission, while recovery WTs correspond to infectious periods or time from a positive test to a negative one. 
Population-level heterogeneity further shapes these distributions. For example, it is expected that elderly, immunocompromised, or high-contact individuals will show longer, more variable recovery and lower shape parameters, whereas children and young adults in community settings will tend to progress and recover more uniformly~\cite{Ali2020,lee2024variability,voinsky2020effects}.

\begin{figure}[ht]
    \centering
    \includegraphics[width=1\linewidth]{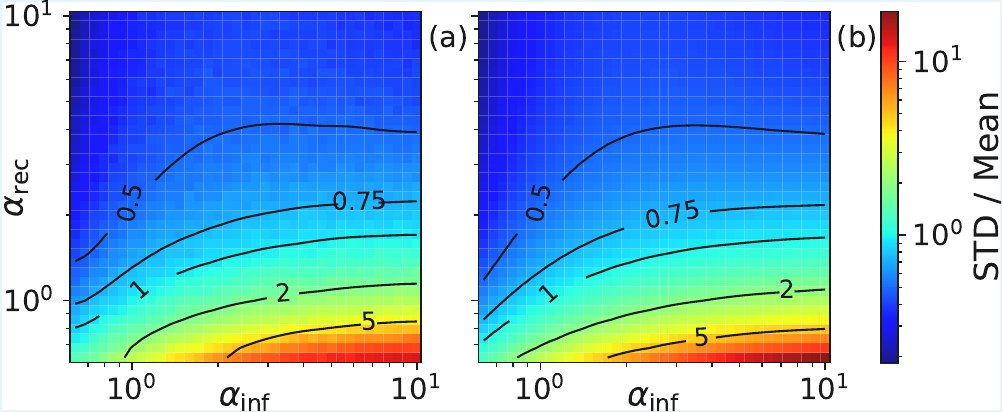}
    \vspace{-9mm}\caption {Numerical (left) and theoretical (right) heatmaps of the outbreak-size distribution COV (normalized by the Markovian COV) versus the infection and recovery shape parameters, for a well-mixed system with $N\!=\!5000$, $R_0\!=\!2$, and $10^3$ runs per point. Black lines are equi-COV contour lines. 
    }
    \label{fig4}
\end{figure}

In Fig.~\ref{fig4} we show the dramatic impact of incorporating non-Markovian reactions. Here  numerical and theoretical heatmaps of the outbreak size distribution's coefficient of variation (COV) are compared for gamma-distributed infection and recovery, versus the shape parameters in a well-mixed setting. Notably, the COV, indicating the relative error in the expected outbreak size, can greatly exceed the Markovian prediction, especially in regions of large infection shape parameter, indicating synchronized and rapid progression from exposure to infectiousness.

Future research should aim at extending our formalism to realistic non-Markovian epidemic settings involving \textit{structured populations} residing on complex heterogeneous networks. \black{The mean-field aspects of non-Markovian epidemics have already been studied on such networks, and the next challenge is to study the interplay between large deviations and complex network topology under non-Markovian dynamics. Therefore, upon successful extension} of the theory to  real-life network topologies, and accounting for empirically grounded shape parameters for both infection and recovery (as done in Fig.~\ref{fig4}), this framework  is expected to provide a more accurate representation of epidemic dynamics, improving estimates of  epidemic thresholds,  outbreak sizes, disease clearance times, and most importantly, intervention measures.

\textit{Acknowledgments}. The authors wish to thank Ami Taitelbaum for many useful discussions.


\bibliography{bibliography.bib}

\vspace{0.5cm}
\LARGE{\textbf{Supplemental Information}}
\normalsize
\setcounter{equation}{0}
\setcounter{figure}{0}
\renewcommand{\theequation}{S\arabic{equation}}
\renewcommand{\thefigure}{S\arabic{figure}}

\section{A. Obtaining the non-Markovian master equation}
We begin by showing how the non-Markovian infection and recovery processes can be formulated. In this aspect, the analysis of the SIS and SIR models are identical, but later the treatment of the master equation will be different depending on the model at hand. We start by computing the survival probabilities corresponding to the waiting-time (WT) distributions between infection and recovery events. \black{For simplicity, we outline the derivation for the case of Markovian infection and non-Markovian recovery for which the WTs are given by Eq.~(\ref{WTrec}) below. Yet, the derivation for the complementary case is identical. The corresponding survival probabilities are}
\begin{equation}
    \Psi_1(t) = e^{-\lambda_1 t} ,\quad \Psi_2(t) = \frac{\Gamma(\alpha,\alpha \lambda_2 t)}{\Gamma(\alpha)},
\end{equation}
\black{where $\Gamma\left(a, z\right)=\int_z^\infty t^{a-1}e^{-t}dt\,$ is the upper incomplete gamma function. }We next calculate $\phi_1(t)$---the distribution for a single infection event to occur at time $t$, assuming no recovery has occurred up to $t$, and $\phi_2(t)$---the  distribution for a single recovery event to occur at time $t$, assuming no infection  has occurred up to $t$. This yields $   \phi_1(t)=\psi_1(t)\Psi_2(t) $, $\phi_2(t)=\psi_2(t)\Psi_1(t)$, and with Eq.~(\ref{WTrec}) below we get
\begin{eqnarray}
    \phi_1(t)&=& NR_0x_sx_i \frac{\Gamma(\alpha,\alpha Nx_i t)}{\Gamma(\alpha)}e^{-NR_0x_sx_i t},\nonumber\\
    \phi_2(t)&=& \frac{(\alpha Nx_i)^\alpha t^{\alpha-1}e^{-\alpha Nx_i t}}{\Gamma(\alpha)} e^{-NR_0x_sx_i t}.
\end{eqnarray}
We then calculate the Laplace transform of $\tilde{\phi}_i$:
\begin{eqnarray}
    \tilde{\phi}_1(u)&=&\frac{NR_0x_sx_i}{u+NR_0x_sx_i}  \newline \left[1\!-\!\left(1\!+\!\frac{u}{\alpha Nx_i}\!+\!\frac{R_0x_s}{\alpha}\right)^{\!\!-\alpha}\right],\nonumber\\
    \tilde{\phi}_2(u)&=&\left(1\!+\!\frac{u}{\alpha Nx_i}\!+\!\frac{R_0x_s}{\alpha}\right)^{\!\!-\alpha}\!,
\end{eqnarray}
where $u$ is the Laplace variable.
Using these quantities, and the framework developed in~[51, 52], we can compute $\Memlap{i}(u)$---the Laplace transform of the memory kernels $\Mem{i}(t)$, which enter the non-Markovian master equation [Eq.~(3) in the main text]---multiplied by the Laplace parameter $u$. This yields
\begin{equation}
\Memlap{i}(u):=u \Lap\left[\Mem{i}(t)\right]=\frac{u \tilde{\phi}_i(u)}{1-\tilde{\phi}_1(u)-\tilde{\phi}_2(u)}.
\label{eq:memlap}
\end{equation}
As stated, the inverse Laplace transform of these memory kernels in Laplace space will enter the non-Markovian master equation, see Eq.~(3) in the main text.
Performing the explicit calculation of these memory kernels using Eq.~(\ref{eq:memlap}), we find
\begin{eqnarray}
    \Memlap{1}(u)&=&N R_0x_ix_s,\nonumber\\
    \Memlap{2}(u)&=&\frac{u+NR_0x_ix_s}{\left[1+u/(\alpha Nx_i)+R_0x_s/\alpha\right]^{\alpha}\!-\!1}.
    \label{Memlap}
\end{eqnarray}
Finally, we use the finite value theorem $\lim_{t\to\infty} f(t)=\lim_{u\to0} u \tilde{f}(u)$, and define the normalized asymptotic memory kernel $\Memf{i}$ as: $
\Memf{i}:=(1/N)\lim_{t\to\infty} \Mem{i}(t)=(1/N)\lim_{u\to0} \Memlap{i}(u)$ to get Eq.~(4) \black{with effective memory kernels: $\Memf{1}=R_0x_ix_s$ and $\Memf{2}=R_0x_ix_s/\left[(1+R_0x_s/\alpha)^{\alpha}-1\right]$. These results complement Eq.~(5) in the main text.}

\section{B. Different choices of waiting-time distributions}
In this section, we show that the asymptotic memory kernels can be obtained in a similar manner also for other choices of WT distributions; for concreteness, we focus on the case of a power-law inter-event time distribution. To simplify matters, we consider here the complementary choice of Markovian infection and non-Markovian recovery, where in Fig.~4 in the main text and in Fig.~\ref{figS3} we show results when both infection and recovery are non-Markovian.

Under exponential infection and power-law distributed recovery,  WT distributions are:
\begin{equation}
    \psi_1(t) = \lambda_1 e^{-\lambda_1 t} ,\quad \psi_2(t) = \frac{\lambda_2 \alpha}{(\alpha-1)(1+\frac{\lambda_2 t}{(\alpha-1)})^{\alpha+1}},
\end{equation}
with means of $\lambda_1=N R_0 x_s x_i$ and $\lambda_2=N x_i$. Here, the shape parameter is bounded, $\alpha>1$, to ensure a finite mean, and in the limit $\alpha\to\infty$ the distribution converges to an exponential one. Notably, the regime of $\alpha>1$ in the gamma distribution does not exist here; namely, by choosing a power-law WT we can only increase the distribution's width compared to an exponential one. 
Performing the explicit calculation of the asymptotic memory kernels, we find 
\begin{equation}\label{MeMPower}
    \Memf{1}=R_0x_ix_s ,\quad \Memf{2}=\frac{\alpha x_i}{\alpha-1}\frac{\Ei{\alpha+1}\left[(\alpha-1)R_0x_s\right]}{\Ei{\alpha}\left[(\alpha-1)R_0x_s\right]},
\end{equation}
where $E_m(z) \equiv \int_1^{\infty}e^{-z\tau }\tau^{-m} d\tau\,$ is the exponential integral function. 

At this point, one can plug these memory kernels [Eqs.~(\ref{MeMGamma}) and (\ref{MeMPower})] into the late-time master equations and perform calculations along the same lines as in the main text, which allows to find the quantities of interest, e.g., outbreak-size distribution within the SIR model, or MTE within the SIS model.

\vspace{0.45cm}
\section{C. Derivation of the final outbreak-size distribution}
To compute the final outbreak size distribution in the realm of the SIR model, we employ the Hamiltonian formalism. Starting from the Hamiltonian
\begin{equation}
\Ham\equiv \Memf{1}(x_s,x_i)\left(e^{p_i-p_s}\!-\!1\right)\!+\!\Memf{2}(x_s,x_i)\left( e^{-p_i}\!-\!1\right),
\label{eq:hamiltonianSM}
\end{equation}
we write down Hamilton's equations 
\begin{eqnarray}
        \hspace{-7mm}\dot x_s&=&\frac{\partial \Ham}{\partial p_s}=-\Memf{1}e^{p_i-p_s},
        \label{eq:x_s}\\
        \hspace{-7mm}\dot x_i&=&\frac{\partial \Ham}{\partial p_i}=\Memf{1}e^{p_i-p_s}-\Memf{2}e^{-p_i},\label{eq:x_i}\\
        \hspace{-7mm}\dot p_s&=&-\frac{\partial \Ham}{\partial x_s}\!=\!\frac{\partial \Memf{1}}{\partial x_s}\left(1\!-\!e^{p_i-p_s}\right)\!+\!\frac{\partial \Memf{2}}{\partial x_s}\left(1\!-\!e^{-p_i}\right)\!,\\
        \hspace{-7mm}\dot p_i&=&-\frac{\partial \Ham}{\partial x_i}\!=\!\frac{\partial \Memf{1}}{\partial x_i}\left(1\!-\!e^{p_i-p_s}\right)\!+\!\frac{\partial \Memf{2}}{\partial x_i}\left(1\!-\!e^{-p_i}\right)\!.
        \label{eq:p_i}
    \end{eqnarray}
The solutions of these equations yield, among other insights, the most probable outbreak dynamics and enables the determination of both the mean outbreak size as well as the full outbreak-size distribution.

Before delving into these equations we point out an important constant of motion, which appears in the SIR model and enables the intrinsically two-dimensional problem to be cast into one dimension. First, we point out that the Hamiltonian is a constant of motion, since the rates do not depend on time explicitly. Furthermore, when examining the rates $\Memf{i}$ we can see that in our example case they are both linear in $x_i$, see Eq.~(5) in the main text. This attribute also holds for a wide variety of other WT distributions such as a power-law distribution, see SI Sec.~B. As a result, using Eqs.~(\ref{eq:hamiltonianSM}) and (\ref{eq:p_i}), we can write the Hamiltonian in a suggestive way $\Ham=-x_i\dot{p_i}$. Finally, since most epidemic waves start from a small fraction of infected, we argue that typically $x_i(t=0)\ll 1$, and is certainly not macroscopic; in fact, it is often assumed that one starts with one infected individual such that $x_i(t=0)\sim O(1/N)$ with $N\gg1$. As a result, since $\Ham(t=0)$ is a constant of motion, one must have $\Ham\simeq0$ throughout the epidemic. Yet, due to the fact that  $\Ham=-x_i\dot{p_i}$, and that $x_i(t)$ changes from a very small value to a macroscopic fraction of the population during the epidemic wave, the only way the Hamiltonian can stay zero is by having $\dot{p_i}=0$ throughout it. Therefore, we have found a new constant of motion: $p_i$. 

Notably, even though in this case $\Ham\simeq 0$, there is an important distinction between our use of the WKB assumption here and in the SIS model. While the condition of $\Ham\simeq0$ is inherent in the SIS model as the distribution becomes metastable with $\partial P(x_i, t)/\partial t\simeq 0$ (up to exponential accuracy), in the SIR model this condition solely stems from starting with a small fraction of infected. 

We now derive the outbreak size distribution by adopting the methodology of~[28] and defining a new constant of motion $m\!=\!e^{p_i}$. Using Eqs.~(\ref{eq:x_s}) and~ (\ref{eq:x_i}) and the fact that the total population is constant, i.e. $\dot{x}_s+\dot{x}_i+\dot{x}_r=0$, we obtain $\dot x_r=-\dot x_s - \dot x_i=\Memf{2}/m$. By equating the Hamiltonian [Eq.~(\ref{eq:hamiltonianSM})] to zero, we can get an expression for $e^{p_s}$
\begin{equation}
          e^{p_s}=\frac{m^2\Memf{1}/\Memf{2}}{m\left(\Memf{1}/\Memf{2}+1\right)-1}.
          \label{eq:exp_p_s}
    \end{equation}
Now we divide $\dot{x}_s$ from Eq.~(\ref{eq:x_s}) by $\dot{x}_r=\Memf{2}/m$ to obtain a differential equation for the dependence of $x_s$ on $x_r$ 
    \begin{equation}
        \frac{dx_s}{dx_r}=-\frac{\Memf{1}}{\Memf{2}}m^2e^{-p_s}=1-m\left(\frac{\Memf{1}}{\Memf{2}}+1\right).
    \end{equation}
By integrating $x_s$ from $1$ to $x_s^*$ and $x_r$ from $0$ to $1-x_s^*$, we obtain an implicit relation between the final outbreak size $x_s^*=x_s(t\to\infty)$ and $m$ 
     \begin{equation}
        \int_1^{x_s^*} \frac{1}{m\left(\Memf{1}/\Memf{2}+1\right)-1}dx_s=\int_0^{1-x_s^*} dx_r.
        \label{eq:m_relation_appendix}
    \end{equation}
Note that, plugging $\alpha=1$ into $\Memf{1}$ and $\Memf{2}$ and using Eq.~(5) in the main text, this result coincides with that of Ref.~[28] in the Markovian case. Equation~(\ref{eq:m_relation_appendix}) is important since it removes $m$ from the action $\Act$  and makes it a function of a single variable $x_s^*$. To find the action $\Act(x_s^*)$ for each final state $x_s^*$, which determines the outbreak size distribution via the WKB ansatz, we write down the formal solution of the action~[25,28]
\begin{equation}
    \Act(x_s,x_i,t)=\int_0^t(p_s\dot{x}_s+p_i\dot{x}_i-\Ham)dt'.
\end{equation}
We argue that the integral over $\Ham$ vanishes because $\Ham\simeq0$, and that when taking $t\to\infty$ the integral over $p_i\dot{x}_i$ vanishes because $p_i$ is constant and $x_i(t=0)=x_i(t\to\infty)\simeq0$. After some algebra, this leaves us with a reduced integral $ \Act(x_s^*)= \int_1^{x_s^*}p_sdx_s$, where $p_s$ is given by Eq.~(\ref{eq:exp_p_s}).
Plugging $p_s$ into the integral, we obtain 
\begin{equation}
    \hspace{-2mm}\Act(x_s^*)=\int_1^{x_s^*}\ln\left\{ m^2\Memf{1}\Big/\left[m\left(\Memf{1}+\Memf{2}\right)-\Memf{2} \right]\right\}dx_s,
        \label{eq:action_sirSM}
\end{equation}
which coincides with Eq.~(8) in the main text. This allows finding the final outbreak-size distribution in the presence of non-Markovian reactions. 

We now provide a more detailed explanation of how to compute the standard deviation. As noted in the main text, we use the Gaussian approximation $\sigma \simeq \left|N \left.\Act''(x_s^*)\right|_{\bar{x}_s^*}\right|^{-1/2}$. Compared to the SIS model, deriving $\Act(x_s^*)$ at $x_s^*=\bar{x}_s^*$ in the SIR model is more challenging for two reasons: the expressions for $\Act(x_s^*)$ and $\bar{x}_s^*$ are more complex, and there is an implicit function $m(x_s^*)$ that must be accounted for in our derivation. To calculate $\left.\Act''(x_s^*)\right|_{\bar{x}_s^*}$ we first need to find $\left.m'(x_s^*)\right|_{\bar{x}_s^*}$ and $\left.m''(x_s^*)\right|_{\bar{x}_s^*}$. Rewriting Eq.~(\ref{eq:m_relation_appendix}) as $F(m, x_s^*)=0$, the derivatives of $m$ follow from the chain rule, $dF/dx_s^*=\left(\partial F/\partial m\right) \left(dm/dx_s^*\right)+\partial F/\partial x_s^*$ with $d F/d x_s^*=0$
\begin{eqnarray}
        \left.m'(x_s^*)\right|_{\bar{x}_s^*}&=&\!\!\left.-\frac{F_{x_s^*}}{F_m}\right|_{\bar{x}_s^*},\\
        \left.m''(x_s^*)\right|_{\bar{x}_s^*}&=&\!\!\left.-\frac{F_m^2F_{{x_s^*}{x_s^*}}\!-\!2F_mF_{x_s^*}F_{{x_s^*}m}\!+\!F_{x_s^*}^2F_{mm}}{F_m^3}\!\right|_{x^*_s=\bar{x}_s^*}\!\!\!\!,\nonumber
    \end{eqnarray}
where $F_z=\partial F/\partial z$. Using these and the implicit formula for $\bar{x}_s^*$ (found by plugging $m(\bar{x}_s^*)=1, x_s^*=\bar{x}_s^*$ in Eq.~(\ref{eq:m_relation_appendix})) we obtain a compact expression for $\left.\Act''(x_s^*)\right|_{\bar{x}_s^*}$
\begin{equation}
   \left.\Act''(x_s^*)\right|_{\bar{x}_s^*} =\frac{\left[\xi(\bar{x}_s^*)-1\right]^2}{1-\bar{x}_s^*+I(\bar{x}_s^*)},
\end{equation}
where $\xi(\bar{x}_s^*)=\Memf{2}(\bar{x}_s^*)/\Memf{1}(\bar{x}_s^*)$, $I(\bar{x}_s^*) = \int_{\bar{x}_s^*}^1\xi(z)^2dz$. For $\alpha=1$, this reduces to $\xi(\bar{x}_s^*)=1/(R_0\bar{x}_s^*)$, $I(\bar{x}_s^*)=-\left(1-\bar{x}_s^*\right)/\left(R_0^2\bar{x}_s^*\right)$, recovering the Markovian result~[28]
\begin{equation}
   \left.\Act''(x_s^*)\right|_{\bar{x}_s^*} =\frac{\left(R_0\bar{x}_s^*-1\right)^2}{\left(1-\bar{x}_s^*\right)\bar{x}_s^*\left(R_0^2\bar{x}_s^*+1\right)}.
\end{equation}

\section{D. Quasi-stationary distribution in the non-Markovian SIS model}

\begin{figure}[ht]
    \centering
    \includegraphics[width=1\linewidth]{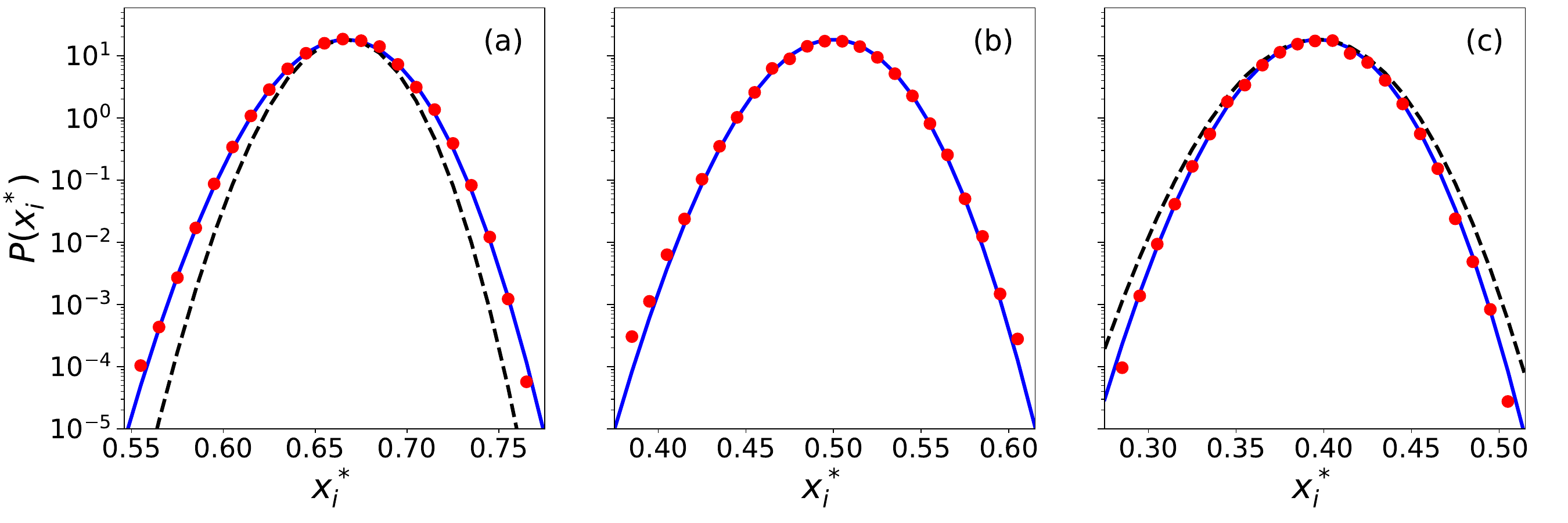}
     \vspace{-8mm} \caption {QSDs at $\alpha\!=\!0.5, 1, 2$ in (a)--(c) in the SIS model of epidemics, for gamma-distributed infection and Markovian recovery, with $N\!=\!1000$, $R_0\!=\!2$,  
     and $10^5$ runs per $\alpha$. Here we compare simulations (symbols) to theory (solid line) and to a Markovian theory adjusted to the same mean (dashed line), with $x_i(0)=x_i^*$. The distributions are computed by averaging over a single trajectory from $t=50$ until $t=10^5$ (in units of $\gamma^{-1}$. 
     }
    \label{figS1}
\end{figure}

Here we compare our analytical prediction for the quasi-stationary distribution (QSD) [Eq.~(13) in the main text] with numerical simulations, for the case of Markovian infection and non-Markovian recovery with gamma-distributed waiting times. In Fig.~\ref{figS1}  we show excellent agreement between the simulated QSDs for various values of $\alpha$ and our theoretical prediction, in the case of non-Markovian, gamma-distributed infection, and Markovian recovery. We also plot by dashed lines the QSDs that would be obtained if we use a Markovian theory with adjusted reaction rates, obtained by replacing the gamma-distributed infection process with an exponential one, and adjusting the infection rate by $R_0\to R_0 \alpha(2^{1/\alpha}-1)$, see main text. While being a good approximation close to the mean, for $\alpha\neq 1$ the tails are missed by this adjusted distribution as can be seen in panels (a) and (c). Notably, this effect strengthens as $\alpha$ is further decreased (or increased). This demonstrates that the non-Markovian effects cannot be fully captured by simply adjusting the mean of a Markovian reaction.

\section{E. SIS model under non-Markovian recovery}
\subsection{E1. SIS model under Markovian infection and Non-Markovian  recovery}
\black{This section is dedicated to the exploration of the SIS model in the complementary case of non-Markovian, gamma-distributed recovery with Markovian infection. The WTs satisfy
\begin{equation}\label{WTrec}
      \psi_1(t) = \lambda_1 e^{- \lambda_1 t},\quad \psi_2(t) = \frac{(\alpha \lambda_2)^\alpha t^{\alpha-1}e^{-\alpha \lambda_2 t}}{\Gamma(\alpha)},
\end{equation}
with means of $\lambda_1=N R_0 x_s x_i$ and $\lambda_2=N x_i$.
Performing the explicit calculation of the normalized asymptotic memory kernels as \black{in Sec.~A}, we find 
\begin{equation}\label{MeMGamma}
\Memf{1}=R_0x_ix_s ,\quad \Memf{2}=\frac{R_0x_ix_s}{\left(1+R_0x_s/\alpha\right)^{\alpha}-1}
\end{equation}
}

\begin{figure}[ht]
    \centering
    \includegraphics[width=1\linewidth]{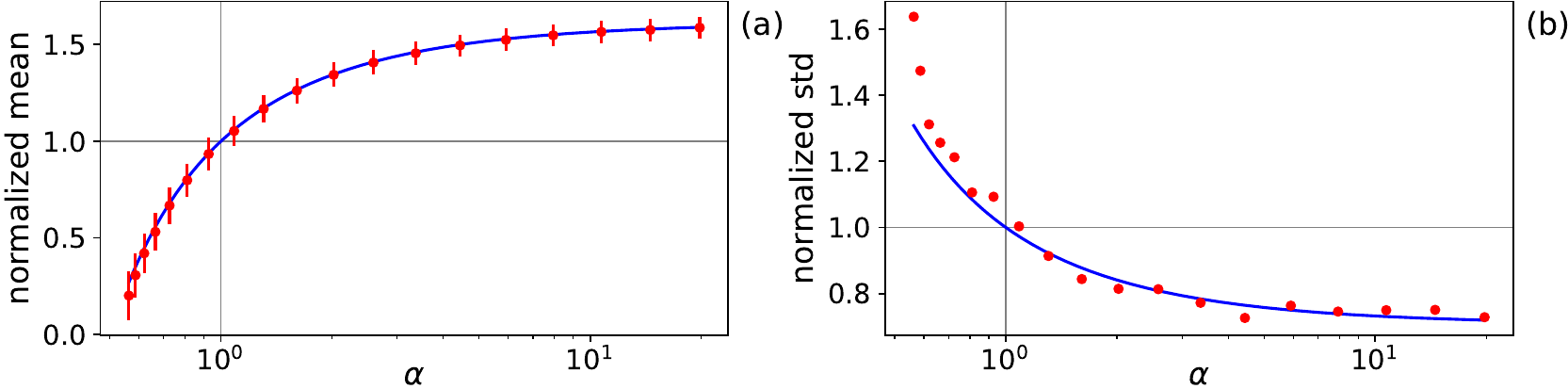}
    \vspace{-7mm}\caption {The normalized mean $x_i^*$  (a) and normalized standard deviation $\sigma$  (b) as functions of $\alpha$, in the SIS model of epidemics, for exponential infection and gamma-distributed recovery, $N=5000$, $R_0=1.5$, and $10^4$ runs  per $\alpha$. Simulations (symbols) are compared with  theory (blue solid line). In all panels the initial condition was $x_i(0)=x_i^*$.}
    \label{figS2}
\end{figure}

\black{Using these memory kernels, we can compute the mean and standard deviation of the QSD, within the SIS model, under Markovian infection and non-Markovian recovery. In Fig.~\ref{figS2}(a,b) we show how the shape parameter $\alpha$ affects the metastable mean and its standard deviation, both normalized by their Markovian values ($x_i^*(1)=1/3$, $\sigma(1)\approx0.026$ for $R_0=1.5$). As $\alpha$ decreases, the mean drastically decreases and the variance increases; this is since the typical time between recovery events gets shorter as the WT distribution becomes wider and more skewed. The opposite occurs for $\alpha>1$; for $\alpha\to\infty$, the mean approaches $x_i^*(\infty)/x_i^*(1)\simeq (1-\ln 2 / R_0)/(1/3)\simeq 1.614$. Near the bifurcation, the mean vanishes and the WKB approximation breaks down.}

\black{The dependence of the variance on $\alpha$ as observed in Fig.~\ref{figS2}(b) can be explained by looking at the right-hand-side of Eq.~(14) in the main text, where the standard deviation $\sigma$ is expressed as a function of the mean $x_i^*$ and $\alpha$ (\black{instead of $R_0$ and $\alpha$}). Similarly as in the case of non-Markovian infection and Markovian recovery, the dominating factor in shaping $\sigma$ is its dependence on $x_i^*$, $\sigma\sim\sqrt{1-x_i^*}$. However, $\sigma$ is also affected by non-Markovianity: $\sigma\sim\left[2\alpha(1-2^{-1/\alpha})\right]^{-1/2}$, which reinforces the dependence on $x_i^*$ in this case. Notably, the WKB approximation breaks down as $x_i^*$ approaches $0$, as seen in the figure.}

\subsection{E2. SIS model under Non-Markovian infection and recovery}
Here, we consider the full non-Markovian case of having both processes of infection and recovery with gamma-distributed WTs. 
%
%
%
In this case the WTs satisfy
\begin{equation}
      \psi_1(t) = \frac{(\alpha \lambda_1)^\alpha t^{\alpha-1}e^{-\alpha \lambda_1 t}}{\Gamma(\alpha)} ,\quad \psi_2(t) = \frac{(\alpha \lambda_2)^\alpha t^{\alpha-1}e^{-\alpha \lambda_2 t}}{\Gamma(\alpha)},
\end{equation}
with means of $\lambda_1=N R_0 x_s x_i$ and $\lambda_2=N x_i$.

\begin{figure}[ht]
    \centering
    \includegraphics[width=1\linewidth]{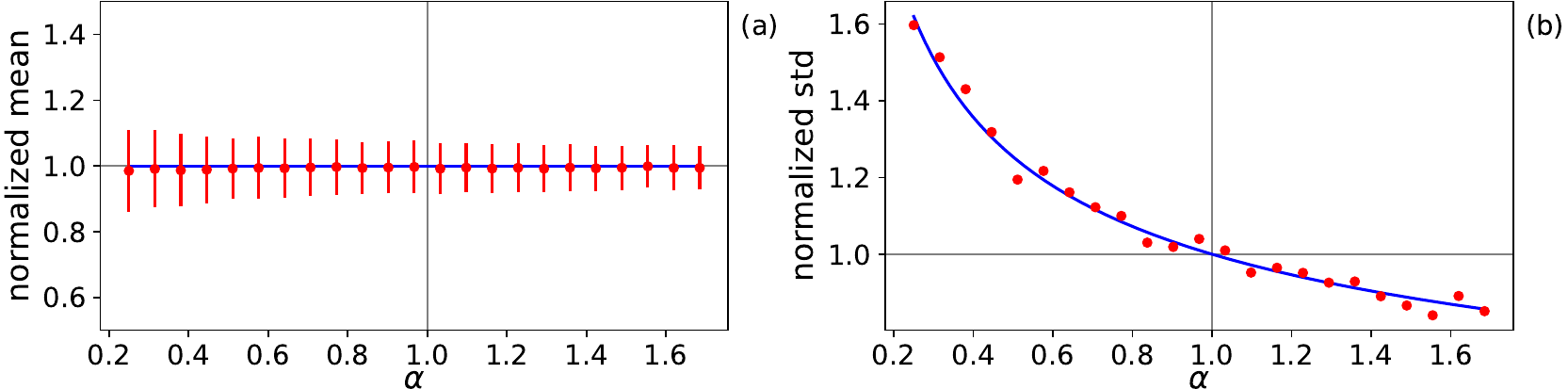}
    \vspace{-7mm}\caption {The normalized mean $x_i^*$  (a) and normalized standard deviation $\sigma$  (b) as functions of $\alpha$, in the SIS model of epidemics, for gamma-distributed infection and gamma-distributed recovery, $N=5000$, $R_0=1.5$, and $10^4$ runs  per $\alpha$. Simulations (symbols) are compared with  theory (blue solid line). In all panels the initial condition was $x_i(0)=x_i^*$.}
    \label{figS3}
\end{figure}

In Fig.~\ref{figS3} we show the dependence of the mean and standard deviation of the QSD on $\alpha$ (where both quantities are normalized as in Fig.~\ref{figS2} and Fig.~3 in the main text), for gamma-distributed infection and recovery with an identical shape parameter. Naturally, this choice is arbitrary; however, while a lengthy analysis can be performed by doing many calculations in the phase space of $(\alpha_1,\alpha_2)$,  we merely wanted to show that an analysis where all the reactions are non-Markovian is feasible. Notably, even a simplified choice of having an identical $\alpha$ value, highlights some of the interesting effects that emerge. As expected, when both processes vary together, the mean remains constant; yet, the standard deviation changes qualitatively in a manner similar to the recovery case (see Fig.~\ref{figS2}). This shows that, although the effects on the mean are exactly inverse for infection and recovery, their impact on fluctuations is not, as is evident from Eq.~(14) in the main text. 

\end{document}